\documentclass[10pt]{article}%
\usepackage{amsmath}
\usepackage{amsfonts}
\usepackage{amssymb}
\usepackage{graphicx}%
\usepackage{xcolor}
\def\nn{\nonumber}

\def\ben{\begin{equation}}
\def\een{\end{equation}}

\def\bea{\begin{eqnarray}}
\def\eea{\end{eqnarray}}

\def\S{\Sigma}

\def\bx{{\bf x}}

\begin{document}

\title{Discrete Newtonian Cosmology: Perturbations}
\author{George F R Ellis,\\ACGC and Department of Mathematics, University of Cape Town,\\Gary W Gibbons\\Trinity College and DAMTP, Cambridge University}
\date{\today }
\maketitle

\begin{abstract}In a previous paper \cite{First} we showed how a finite system of discrete particles interacting with each other via Newtonian gravitational attraction would lead to precisely the same dynamical equations for homothetic motion as in the case of the pressure-free Friedmann-Lema\^{i}tre-Robertson-Walker cosmological models of General Relativity Theory, provided the distribution of particles obeys the central configuration equation. In this paper we show one can obtain perturbed such Newtonian solutions that give the same linearised structure growth equations as in the general relativity case. We also obtain the  Dmitriev-Zeldovich equations for subsystems in this discrete gravitational model, and show how it leads to the conclusion that voids have an apparent negative mass.
\end{abstract}

\section{Introduction}

\label{sec:intro} This is the second part of a treatment of Discrete Newtonian
Cosmology based on a point particle model according to  which, in contrast to the usual
fluid models, the universe is conceived of as consisting of a large number $N$
of gravitating point particles of mass $m_{a}$ and positions $\mathbf{x}%
_{a}(t)$ acted upon by Newtonian gravity and a possible cosmological term. In
our first paper \cite{First} we laid down the foundations and described how
homothetic solutions $\mathbf{x}_{a}=S(t)\mathbf{r}_{a}$ may be constructed
which are the analogues of the Friedmann-Lema\^{i}tre models of the continuum
theory. The scale factor $S(t)$ was shown to exactly satisfy the Raychaudhuri
equation of gravitational attraction provided the co-moving positions
$\mathbf{r}_{a}$ constitute a \textit{central configuration }(see\textit{
}(\ref{Newt3})). In previous work \cite{BatGibSut03} it has been shown that
for $N$ large and all masses $m_{a}$ equal , there exist central
configurations for which the point particles are distributed in an extremely
homogeneous and isotropic fashion within a ball of finite radius. Thus one
obtains the same results as in the fluid case, but without making the fluid
assumption, which is somewhat dubious in this context \cite{First}.
After all most of the material content of the universe 
appears to be in the form of cold dark matter whose precise nature 
is unknown except that it probably consists of a non-interacting
gas of particles which interact solely by gravitational forces.
In our Newtonian model we need only assume that the dominant material content of 
the universe consists of particles moving non-relativistically
 whose masses we need not specify
and which interact solely by  Newton's inverse square law of gravitation.
\bigskip

In this paper we investigate the behaviour of inhomogeneous discrete Newtonian
cosmological models representing small deviations from that cosmological
background. After reviewing the basic theory and the exact homothetic
solutions in we shall, in Section \ref{sec:pert}, outline how perturbations
around a general solution of Newtons' equations of motion behave. We then
apply this general theory to homothetic solutions, obtaining the discrete
Newtonian analogue of perturbed relativistic cosmological models. This gives
the same equations of motion as fluid-based Newtonian perturbation theory
\cite{Ell90} , which is also the same as in the pressure-free
General\ Relativity case \cite{EllBru89} .
 We go on in Section \ \ref{sec:DZ}
to derive, following \cite{GibPat03}, what we call the
\textit{Dmitriev-Zeldovich equations}. This is a rather different approach to
perturbation theory \cite{Zel}, in which we obtain equations governing the
motion of Newtonian point particles in a background Friedmann-Lemaitre
cosmology. This is a mean-field theory in which the point particles interact
gravitationally with each other but have negligible effect on the background.
The resulting equations are widely used in investigations of large-scale
structure in cosmology \cite{BagPad97,Ber98}.In Section 5, we relate this to the Swiss Cheese approximation used in General Relativity, and comment on  the apparent negative mass of voids, in accordance with Newtonian work by F\"{o}ppl and general relativity comments by Bondi.
\\

%{\color{blue}** add re final sections**}\\

%\subsection{Basic Theory}

In the remainder of this section we summarise the discrete Newtonian theory that was set out in
\cite{First}, giving the general exact dynamic equations, plus the exact
homothetic solution for the background cosmology.

\subsection{Equations of motion}

Consider an isolated set of gravitating particles, with no other interparticle
forces. The gravitational force of the $b$-th particle on the $a$-th particle
is
\begin{equation}
\mathbf{F}_{ab}=-\frac{Gm_{a}m_{b}}{|\mathbf{x}_{a}-\mathbf{x}_{b}|^{3}%
}(\mathbf{x}_{a}-\mathbf{x}_{b})\,=-\mathbf{F}_{ba}, \label{grav}%
\end{equation}
where $G$ is Newton's gravitational constant. The equation of motion for the
$a-$th particle is
\begin{equation}
m_{a}\frac{d^{2}\mathbf{x}_{a}}{dt^{2}}=-\sum_{b\neq a}Gm_{a}m_{b}%
\frac{(\mathbf{x}_{a}-\mathbf{x}_{b})}{|\mathbf{x}_{a}-\mathbf{x}_{b}|^{3}%
}=\mathbf{F}_{a} \label{Newt}%
\end{equation}
where $\mathbf{F}_{a}$ is the total gravitational force acting on the $a$-th
particle due to all the other particles in the system. It can be represented
in terms of the gravitational potential energy $V_{a}$ of the particle $a$ due
to all the other particles, defined by
\begin{equation}
V_{a}(\mathbf{x}_{a}):=-\sum_{b\neq a}\frac{Gm_{a}m_{b}}{|\mathbf{x}%
_{a}-\mathbf{x}_{b}|}.
\end{equation}
(this clearly depends on the position of the particle $a$). The gravitational
force on the $a$-th particle due to the system of particles is the gradient of
this potential:
\begin{equation}
\frac{\partial V_{a}}{\partial\mathbf{x}_{a}}=-\sum_{b\neq a}\mathbf{F}%
_{ab}=-\mathbf{F}_{a}.
\end{equation}
%Summing over $a$, \ the skew symmetry of $\mathbf{F}_{ab}$ gives
%\begin{equation}
%\sum_{a}\frac{\partial V_{a}}{\partial\mathbf{x}_{a}}=-\sum_{a}\sum_{b\neq
%a}\mathbf{F}_{ab}=0.
%\end{equation}
%which is the discrete version of Laplace's equation (there is no mass in the
%empty space between the particles, so the divergence of the potential vanishes).

Because particle mass $m_{a}$ is conserved,the equations are invariant under
time reversal, time translations, spatial translations, and rotations. In
accordance with Noether's theorem, there are conserved quantities associated
with each of the three continuous symmetries. In particular, total energy
$\mathcal{E}$ of the set of particles is conserved:
\begin{equation}
\mathcal{E}=T+V=\mathcal{E}_{0}\,(\mathrm{constant}),\;
\label{eq:energy_const}%
\end{equation}
where the total kinetic energy $T(\dot{\mathbf{x}_{1}},\dot{\mathbf{x}_{2}%
},\dots,\dot{\mathbf{x}}_{N})$ and the total potential energy $\,V(\mathbf{x}%
_{1},\mathbf{x}_{2}\dots,\mathbf{x}_{N})\,\ $ are defined by%

\begin{equation}
T(\dot{\mathbf{x}} _{1} ,\dot{\mathbf{x}} _{2} , \dots, \dot{\mathbf{x}} _{N}
) := \frac{1}{2} \sum_{a} m_{a} ( \dot{\mathbf{x}} _{a} )^{2}%
\end{equation}

\begin{equation}
V(\mathbf{x}_{1}, \mathbf{x}_{2} \dots, \mathbf{x}_{N} ) = \frac{1}{2}
\sum_{a} V_{a} =-\sum_{1\le a \le b \le N} \frac{Gm_{a}m_{b}}{|\mathbf{x}%
_{a}-\mathbf{x}_{b}|}. \label{potential}%
\end{equation}
These are just single numbers for the entire set of particles:\ coarse-grained
representations of its total internal state of motion and its total
gravitational self-interaction. Thus neither is a function of position.

\subsection{Homothetic ansatz}

To obtain the background cosmological model, we assume self-similarity of the
solution \cite{First}. Then there is a homothetic factor $S(t)$ such that
\begin{equation}
\mathbf{x}_{a}=S(t)\mathbf{r}_{a},\,\,d\mathbf{r}_{a}/dt=0, \label{homothetic}%
\end{equation}
where $\mathbf{r}_{a}$ are co-moving coordinates for the particle $a$. The
total mass of matter $M$\ in a co-moving volume $V$ is given by $M_{V}:=%
%TCIMACRO{\dsum \limits_{a\in V}}%
%BeginExpansion
{\displaystyle\sum\limits_{a\in V}}
%EndExpansion
m_{a}$ which is conserved. The volume scales as $V=S^{3}(t)V_{0}$ so the
density scales as%

\begin{equation}
\rho:=\frac{M_{V}}{V}=\frac{M_{V}}{S^{3}(t)V_{0}}=\frac{\rho_{0}}{S^{3}(t)\ },
\end{equation}
where $\rho_{0}:=\frac{M_{V}}{V_{0}}.$

Define $C(t):=S^{2}(t)\frac{d^{2}S(t)}{dt^{2}}$ and substitute into the
equation of motion (\ref{Newt}); then consistency demands that
$C(t)=\mathrm{const}=:-G\tilde{M},$ where $\tilde{M}$ is the effective
gravitational mass of the system, and the equation separates into the
\emph{central configuration equation}
\begin{equation}
\tilde{M}m_{a}\mathbf{r}_{a}=\sum_{b\neq a}m_{a}m_{b}\frac{(\mathbf{r}%
_{a}-\mathbf{r}_{b})}{|\mathbf{r}_{a}-\mathbf{r}_{b}|^{3}}\label{Newt3}%
\end{equation}
which must hold for all values $a$ (\cite{BatGibSut03}; \cite{Arnold}:79-80),
which is a consistency condition for (\ref{homothetic}) to give a solution,
and the Raychaudhuri equation
\begin{equation}
-\frac{G\tilde{M}}{S^{2}(t)}=\frac{d^{2}S(t)}{dt^{2}}\label{eq:ray}%
\end{equation}
which gives the time evolution. Equation (\ref{Newt3}) determines the value of
$\tilde{M},$ which is not the same as $M_{V}.$ Defining the effective
potential
\begin{equation}
\tilde{V}_{(-1)}:=-\sum_{1\leq\ <b\leq N}\frac{Gm_{a}m_{b}}{|\mathbf{r}%
_{a}-\mathbf{r}_{b}|}%
\end{equation}
of the total system of particles and its effective moment of inertia
\begin{equation}
\tilde{I}_{0}:=\frac{1}{2}\sum_{a}m_{a}(\mathbf{r}_{a})^{2}%
\end{equation}
in terms of the co moving $\mathbf{r}_{a}$, these are both constants. A key
identity following from the central configuration equation is
\begin{equation}
2G\tilde{M}\,\tilde{I}_{0}=\ -\tilde{V}_{(-1)},\;\,\label{V0}%
\end{equation}
which can be used to determine $\tilde{M}.$ In consequence of this identity,
the energy conservation equation(\ref{eq:energy_const})\ is equivalent to the
usual Friedmann equation
\begin{equation}
\frac{1}{2}\left[  \frac{\dot{S}(t)}{S(t)}\right]  ^{2}=\frac{G\tilde{M}%
}{S^{3}(t)}+\frac{E}{S^{2}(t)}\label{Fried12}%
\end{equation}
for pressure-free matter, where $E:=\frac{\mathcal{E}_{0}}{2\tilde{I}_{0}}$ is
a rescaled version of the total internal energy of the system, see
(\ref{eq:energy_const}). This is a first integral of the Raychaudhuri equation
(\ref{eq:ray}) .

\section{Perturbations}

\label{sec:pert}In this section first we perturb the generic equations, and
then apply that method to obtain a perturbed form of the homothetic solutions.

\subsection{The general case}

The general form of the equations of motion we consider is
\begin{equation}
m_{a}\ddot{{\mathbf{x}}}_{a}=-\frac{\partial V({\mathbf{x}}_{1},{\mathbf{x}%
}_{2},\dots{\mathbf{x}}_{N})}{\partial{\mathbf{x}}_{a}}\,.
\end{equation}
where $V({\mathbf{x}}_{1},{\mathbf{x}}_{2},\dots{\mathbf{x}}_{N})$ is the
mutual gravitational potential energy of our $N$ particles, given by
(\ref{potential}) . Actually this is a master equation that applies for any
conservative kind of force; our specific application is where only
gravitational forces act.

\subsubsection{Potential form and Hessian}

Now consider a background solution given by $\bar{{\mathbf{x}}}_{a}$ and
linear perturbation $\delta{\mathbf{y}}_{a}$ about this solution, so that%

\begin{equation}
{\mathbf{x}}_{a}=\bar{{\mathbf{x}}}_{a}+\delta{\mathbf{y}}_{a},\;|\bar
{{\mathbf{x}}}_{a}|\gg|\delta{\mathbf{y}}_{a}|.
\end{equation}
A simple use of Taylor's theorem, neglecting second order terms in
$\delta{\mathbf{y}}_{a}$ yields%

\begin{eqnarray}
m_{a}\left[  \frac{d^{2}(\bar{{\mathbf{x}}}_{a})}{dt^{2}}+\frac{d^{2}%
(\delta{\mathbf{y}}_{a})}{dt^{2}}\right]   & =&m_{a}\frac{d^{2}(\bar
{{\mathbf{x}}}_{a}+\delta{\mathbf{y}}_{a})}{dt^{2}}\nn\\
& =&-\frac{\partial V(\bar{{\mathbf{x}}}_{a}+\delta{\mathbf{y}}_{a})}%
{\partial{\mathbf{x}}_{a}}\nn\\ &=&-
\left[  \frac{\partial V(\bar{{\mathbf{x}}}_{a}%
)}{\partial{\mathbf{x}}_{a}}+\frac{\partial^{2}V(\bar{{\mathbf{x}}}_{a}%
)}{\partial{\mathbf{x}}_{a}\partial{\mathbf{x}}_{b}}\centerdot\,\partial
{\mathbf{x}}_{b}\right]  .
\end{eqnarray}
Cancelling the background terms, the perturbation equation is
\begin{equation}
m_{a}\ddot{\delta{\mathbf{y}}}_{a}=-\sum_{b\neq a}\frac{\partial^{2}%
V}{\partial{\mathbf{x}}_{a}{\partial{\mathbf{x}}_{b}}}(\bar{{\mathbf{x}}}%
_{1},\bar{{\mathbf{x}}}_{2},\dots\bar{{\mathbf{x}}}_{N})\centerdot
{\delta{\mathbf{y}}}_{b}\,. \label{Hess}%
\end{equation}
The symmetric linear operator acting on ${\mathbf{y}}_{a}$ is in
fact minus the Hessian ${\mathbf{E}}_{ab}$ of $V$, considered as a function on
the 3N-dimensional configuration space evaluated on the background solution:
\begin{equation}
m_{a}\ddot{\delta{\mathbf{y}}}_{a}=\sum_{b\neq a}{\mathbf{E}}_{ab}%
.{\delta{\mathbf{y}}}_{b},\,\;{\mathbf{E}}_{ab}:=-\frac{\partial^{2}V_{a}%
(\bar{{\mathbf{x}}}_{c})}{\partial{\mathbf{x}}_{a}{\partial{\mathbf{x}}_{b}}}.
\label{Hess2} \end{equation}

In general (\ref{Hess}) or equivalently (\ref{Hess2}) 
is a linear ordinary differential equation for the perturbation 
$\delta \mathbf{y}_a(t)$  whose  coefficients depend on the background solution
${\bar {\mathbf{x}} }  _a(t)$. These coefficients will in general therefore be 
time dependent.
Equation (\ref{Hess}) was obtained in
the case of 4 particles undergoing a homothetic motion in \cite{Whiting}, and
an evaluation of the resulting Hessian carried out.

\subsubsection{Force form}

Using the expression $\mathbf{F}_{ab}=-\frac{Gm_{a}m_{b}}{|\mathbf{x}%
_{a}-\mathbf{x}_{b}|^{3}}(\mathbf{x}_{a}-\mathbf{x}_{b})$\ for the force
between the particles at ${\mathbf{x}}_{a}$ and ${\mathbf{x}}_{b}$ and setting
$\mathbf{x}_{ba}:=\mathbf{x}_{b}-\mathbf{x}_{a},\,\delta{\mathbf{y}}%
_{ba}=\delta{\mathbf{y}}_{b}-\delta{\mathbf{y}}_{a},x_{ba}:=|\mathbf{x}%
_{b}-\mathbf{x}_{a}|=\left(  (\mathbf{x}_{b}-\mathbf{x}_{a}).(\mathbf{x}%
_{b}-\mathbf{x}_{a})\right)  ^{1/2}$ gives%

\begin{align*}
\mathbf{F}_{ab}(\bar{{\mathbf{x}}}_{a}+\delta{\mathbf{y}}_{a})  &
=-\frac{Gm_{a}m_{b}}{|({\mathbf{\bar{x}}}_{b}-{\mathbf{\bar{x}}}_{a}%
)+\delta{\mathbf{y}}_{ab}|^{3}}(({\mathbf{\bar{x}}}_{a}-{\mathbf{\bar{x}}}%
_{b})\ +\delta{\mathbf{y}}_{ab})\\
&  =-\frac{Gm_{a}m_{b}}{|{\mathbf{\bar{x}}}_{b}-{\mathbf{\bar{x}}}_{a}|^{3}%
}(\bar{{\mathbf{x}}}_{ab}+\delta{\mathbf{y}}_{ab})-\frac{\partial}%
{\partial\mathbf{x}_{a}}\left[  \frac{Gm_{a}m_{b}}{|\mathbf{x}_{b}%
-\mathbf{x}_{a}|^{3}}\right]  \centerdot\delta{\mathbf{y}}_{ab}\mathbf{\ }%
{\mathbf{x}}_{ab}\;\mathbf{+\;O(}\delta{\mathbf{y}}_{a})^{2}%
\end{align*}
to first order, where the partial derivative $(\partial/\partial\mathbf{x}%
_{a})$ is taken keeping all the other positions $\mathbf{x}_{b}\,\,(b\neq a)$
constant. For $\mathbf{x}_{a}\neq\mathbf{x}_{b}$,
\[
\frac{\partial}{\partial\mathbf{x}_{a}}(x_{ba})=-\frac{1}{2}\left(
(\mathbf{x}_{b}-\mathbf{x}_{a}).(\mathbf{x}_{b}-\mathbf{x}_{a})\right)
^{-1/2}2(\mathbf{x}_{b}-\mathbf{x}_{a})=-\left(  x_{ba}\right)  ^{-1}%
\mathbf{x}_{ba}%
\]
This gives
\begin{align}
\delta{\mathbf{F}}_{ab}  &  =\mathbf{F}_{ab}(\bar{{\mathbf{x}}}_{a}%
+\delta{\mathbf{y}}_{a})-\mathbf{F}_{ab}(\bar{{\mathbf{x}}}_{a})\nonumber\\
&  =\frac{Gm_{a}m_{b}}{|\bar{{\mathbf{x}}}_{ab}|^{3}}\delta{\mathbf{y}}%
_{ba}-3\frac{Gm_{a}m_{b}}{|\bar{{\mathbf{x}}}_{ab}|^{4}}(\frac{\partial
}{\partial\mathbf{x}_{a}}(x_{ba})\centerdot\delta{\mathbf{y}}_{ab}%
){\mathbf{\bar{x}}}_{ba}\\
&  =\frac{Gm_{a}m_{b}}{|\bar{{\mathbf{x}}}_{ab}|^{5}}\left\{  \delta
{\mathbf{y}}_{ba}|{\mathbf{\bar{x}}}_{ab}|^{2}-3({\mathbf{\bar{x}}}%
_{ba}\centerdot\delta{\mathbf{y}}_{ba}){\mathbf{\bar{x}}}_{ba}\right\}
\end{align}
and so
\begin{equation}
m_{a}(\delta{\mathbf{y}}_{a})\ddot{}=\sum_{b\neq a}\frac{Gm_{a}m_{b}%
}{{\mathbf{\bar{x}}}_{ab}{}^{5}}\left\{  \delta{\mathbf{y}}_{ba}%
{\mathbf{\bar{x}}}_{ab}^{2}-3({\mathbf{\bar{x}}}_{ba}\centerdot\delta
{\mathbf{y}}_{ba}){\mathbf{\bar{x}}}_{ba}\right\}  \label{pert3}%
\end{equation}

This applies generically to perturbations about any background.

\subsection{The cosmology case}

We now apply the general formalism to the homothetically expanding background
solution described in Section 1. Thus we have
\begin{equation}
{\mathbf{\bar{x}}}_{a}=S(t){\mathbf{\bar{r}}}_{a},\ {\mathbf{\bar{r}}}%
_{a}=const,,{\mathbf{\bar{r}}}_{ab}:={\mathbf{\bar{r}}}_{a}-{\mathbf{\bar{r}}%
}_{b}=const,\,\;\,\bar{r}_{ab}:=|{\mathbf{\bar{r}}}_{a}-{\mathbf{\bar{r}}}%
_{b}|=const. \label{pertc1}%
\end{equation}
Define co moving perturbation variables ${\mathbf{S}}_{a}(t),$ ${\mathbf{S}%
}_{ba}(t)$ by
\begin{equation}
\delta{\mathbf{y}}_{a}=S(t){\mathbf{S}}_{a}(t),\;{\mathbf{S}}_{ba}%
(t):={\mathbf{S}}_{b}-{\mathbf{S}}_{a}.
\end{equation}
Then eqn (\ref{pert3}) becomes%
\begin{eqnarray}
m_{a}\frac{d^{2}}{dt^{2}}(S(t){\mathbf{S}}_{a})&=&\sum_{b\neq a}\frac
{Gm_{a}m_{b}}{S^{5}(t)|{\mathbf{\bar{r}}}_{a}-{\mathbf{\bar{r}}}_{b}|^{5}
}S^{3}(t)\{  ({\mathbf{S}}_{b}-{\mathbf{S}}_{a})|{\mathbf{\bar{r}}}
_{a}-{\mathbf{\bar{r}}}_{b}|^{2} \nonumber
\\ 
&-& 3({\mathbf{\bar{r}}}_{b}-{\mathbf{\bar{r}}
}_{a})\centerdot\mathbf{S}_{ba})({\mathbf{\bar{r}}}_{b}-{\mathbf{\bar{r}}}
_{a})\} \,,
\end{eqnarray}
giving the cosmological perturbation equation
\begin{equation}
S^{2}m_{a}\frac{d^{2}}{dt^{2}}(S{\mathbf{S}}_{a})=\sum_{b\neq a}\frac
{Gm_{a}m_{b}}{|{\mathbf{\bar{r}}}_{a}-{\mathbf{\bar{r}}}_{b}|^{5}}\left\{
\bar{r}_{ba}^{2}{\mathbf{S}}_{ba}-3({\mathbf{\bar{r}}}_{ba}\centerdot
{\mathbf{S}}_{ba}){\mathbf{\bar{r}}}_{ba}\right\}  \label{pertc3}
\end{equation}

As in the general case discussed earlier  (\ref{pertc3}) 
is a second order ordinary differential equation for the 
perturbation $\mathbf{S}_a (t)$ whose coefficients
depend upon the background scale factor $S(t)$ and the background
time independent central configuration $ \mathbf{\bar r} _a $ 
whose homethetic expansion we are  perturbing about.
Since we are not changing the masses $m_a$  in the 
central configuration equation
its solutions, which are critical points of a fixed function on 
configuration space,  will generically be isolated,  
and so in fact there are no 
static small perturbations of the central configuration equation
to consider.

\subsubsection{Asymptotic solution}

Multiply by $(1/m_{a}S^{2})$, the growth of perturbations is given by
\begin{equation}
\frac{d^{2}}{dt^{2}}(S\,{\mathbf{S}}_{a})=\frac{1}{S^{2}}\sum_{b\neq c}%
\frac{Gm_{b}}{\bar{r}_{ab}^{5}}\left\{  \bar{r}_{ba}^{2}{\mathbf{S}}%
_{ba}-3({\mathbf{\bar{r}}}_{ba}\centerdot{\mathbf{S}}_{ba}){\mathbf{\bar{r}}%
}_{ba}\right\}  \label{pertc4}%
\end{equation}
The right hand side goes to zero as $S\rightarrow\infty$. Thus at late times
\begin{equation}
S\,{\mathbf{S}}_{a}=\mathbf{w}_{a}t+{\mathbf{q}}_{a} \label{pertgrow}%
\end{equation}
where $\mathbf{w}_{a},{\mathbf{q}}_{a}$ are constant vectors, and so, because
$S\propto t^{2/3},$
\begin{equation}
{\mathbf{S}}_{a}={\mathbf{\tilde{w}}}_{a}t^{1/3}+\frac{{\mathbf{\tilde{q}}%
}_{a}}{t^{2/3}}. \label{eq:pertgrow1}%
\end{equation}
The first term grows only algebraically, while the second term decays, so
eventually ${\mathbf{S}}_{a}\propto t^{1/3}$. The magnitude of the change is%

\begin{equation}
{\mathbf{S}}^{2}={\mathbf{S}}_{a}{\mathbf{S}}_{a}=\left(  {\mathbf{\tilde{w}}%
}_{a}t^{1/3}+\frac{{\mathbf{\tilde{q}}}_{a}}{t^{2/3}}\right)  \left(
{\mathbf{\tilde{w}}}_{a}t^{1/3}+\frac{{\mathbf{\tilde{q}}}_{a}}{t^{2/3}%
}\right)  .
\end{equation}
so at late times ${\mathbf{S}}^{2}={\mathbf{\tilde{w}}}^{2}t^{2/3}\mathbf{.}$

\subsection{The density perturbation equation}

The mass of matter $M$\ in a co moving volume $V$ is given by $M_{V}:=%
{\displaystyle\sum\limits_{a\in V}}
m_{a},$ which is conserved when the system is perturbed (particle mass is
unchanged). But then $V=S^{3}(t)(V_{0}+\delta V)$ where $\delta V\ \ $is found
by choosing three vectors $\mathbf{x}_{ab}^{\mathbf{i}},\mathbf{x}%
_{ac}^{\mathbf{j}},\mathbf{x}_{ad}^{\mathbf{k}}$ linking particle $a$ to
particles $b,c,d$. The volume defined by these particles is%

\begin{align*}
V_{abcd}  &  =\varepsilon_{ij\bar{k}}\mathbf{x}_{ab}^{\mathbf{i}}%
\mathbf{x}_{ac}^{\mathbf{j}}\mathbf{x}_{ad}^{\mathbf{k}}=\varepsilon
_{ij\bar{k}}(\bar{\mathbf{x}}_{ab}^{i}+\delta\mathbf{y}_{ab}^{\mathbf{i}%
})(\bar{\mathbf{x}}_{ac}^{j}+\delta\mathbf{y}_{ac}^{\mathbf{j}})(\bar
{\mathbf{x}}_{ad}^{k}+\delta\mathbf{y}_{ad}^{\mathbf{k}})\\
&  =\bar{V}_{abcd}+\ \varepsilon_{ij\bar{k}}(\bar{\mathbf{x}}_{ac}^{j}%
\bar{\mathbf{x}}_{ad}^{k}\delta\mathbf{y}_{ab}^{\mathbf{i}}+\bar{\mathbf{x}%
}_{ad}^{k}\bar{\mathbf{x}}_{ab}^{i}\delta\mathbf{y}_{ac}^{\mathbf{j}}%
+\bar{\mathbf{x}}_{ab}^{i}\bar{\mathbf{x}}_{ac}^{j}\delta\mathbf{y}%
_{ad}^{\mathbf{k}})+o(\delta^{2})
\end{align*}
In the cosmological case this is%
\[
V_{abcd}=\bar{V}_{abcd}+\ S^{3}(t)\ \varepsilon_{ij\bar{k}}(\bar{\mathbf{r}%
}_{ac}^{j}\bar{\mathbf{r}}_{ad}^{k}S_{ab}^{i}+\bar{\mathbf{r}}_{ac}^{j}%
\bar{\mathbf{r}}_{ab}^{i}S_{ac}^{j}+\bar{\mathbf{r}}_{ab}^{i}\bar{\mathbf{r}%
}_{ac}^{j}S_{ad}^{k})
\]

At late times they obey (\ref{pertgrow}) so the volume $\delta V$ behaves as%

\begin{align*}
\delta V  &  =S^{3}(t)\ \varepsilon_{ij\bar{k}}(\bar{\mathbf{r}}_{ac}^{j}%
\bar{\mathbf{r}}_{ad}^{k}S_{ab}^{i}+\bar{\mathbf{r}}_{ac}^{j}\bar{\mathbf{r}%
}_{ab}^{i}S_{ac}^{j}+\bar{\mathbf{r}}_{ab}^{i}\bar{\mathbf{r}}_{ac}^{j}%
S_{ad}^{k}),\\
\;{\mathbf{S}}_{ab}^{i}  &  :=({\mathbf{w}}_{a}^{i}t+{\mathbf{q}}%
_{a}^{{\mathbf{i}}})-({\mathbf{w}}_{b}^{i}t+{\mathbf{q}}_{b}^{{\mathbf{i}}%
})=({\mathbf{w}}_{a}^{i}-{\mathbf{w}}_{b}^{i})t+({\mathbf{q}}_{a}%
^{{\mathbf{i}}}-{\mathbf{q}}_{b}^{{\mathbf{i}}})
\end{align*}
Thus\emph{\textbf{ }}their density changes as
\[
\rho:=\frac{M}{(V+\delta V)}\approx\frac{M}{S^{3}(t)\ }(1-\varepsilon
_{ij\bar{k}}(\bar{\mathbf{r}}_{ac}^{j}\bar{\mathbf{r}}_{ad}^{k}S_{ab}^{i}%
+\bar{\mathbf{r}}_{ac}^{j}\bar{\mathbf{r}}_{ab}^{i}S_{ac}^{j}+\bar{\mathbf{r}%
}_{ab}^{i}\bar{\mathbf{r}}_{ac}^{j}S_{ad}^{k})=\rho+\delta\rho,
\]
So finally density perturbations overall for large $\ t$ are given by%

\begin{eqnarray}
\frac{\delta\rho}{\rho}&=&-\frac{1}{n}\sum_{a,b,c,d}\varepsilon_{ij\bar{k}}
(\bar{\mathbf{r}}_{ac}^{j}\bar{\mathbf{r}}_{ad}^{k}({\mathbf{w}}_{a}
^{i}-{\mathbf{w}}_{b}^{i})t\nonumber \\ &+&
\bar{\mathbf{r}}_{ac}^{j}\bar{\mathbf{r}}_{ab}
^{i}({\mathbf{w}}_{a}^{i}-{\mathbf{w}}_{c}^{i})t+\bar{\mathbf{r}}_{ab}^{i}
\bar{\mathbf{r}}_{ac}^{j}({\mathbf{w}}_{a}^{i}-{\mathbf{w}}_{d}^{i})t) \nonumber \\ &\approx&
Wt\propto S^{3/2}
\end{eqnarray}

$W$ depends on initial conditions. If $W>0$ we have the 
growth of an over-density, if $W<0$  the growth of an under density or void. 

\section{Cosmological constant}
\label{cc} The universe appears today to be dominated by a cosmological
constant. Adding in a Newtonian cosmological constant to the force law, we
get
\begin{equation}
m_{a}\frac{d^{2}\mathbf{x}_{a}}{dt^{2}}=-\sum_{b\neq a}Gm_{a}m_{b}%
\frac{(\mathbf{x}_{a}-\mathbf{x}_{b})}{|\mathbf{x}_{a}-\mathbf{x}_{b}|^{3}%
}+\frac{\Lambda m_{a}\mathbf{x}_{a}}{3}. \label{Newtlamb}%
\end{equation}

\subsubsection{Perturbations with cosmological constant}

\bigskip Now consider a background solution given by $\bar{{\mathbf{x}}}_{a}$
and linear perturbation $\delta{\mathbf{y}}_{a}$ about this solution, so that
as before, ${\mathbf{x}}_{a}=\bar{{\mathbf{x}}}_{a}+\delta{\mathbf{y}}%
_{a},\;|\bar{{\mathbf{x}}}_{a}|\gg|\delta{\mathbf{y}}_{a}|.$Again, a simple
use of Taylor's theorem, neglecting second terms in $\delta{\mathbf{y}}_{a}$ yields%

\begin{align}
m_{a}\frac{d^{2}(\bar{{\mathbf{x}}}_{a}+\delta{\mathbf{y}}_{a})}{dt^{2}}  &
=m_{a}\left[  \frac{d^{2}(\bar{{\mathbf{x}}}_{a})}{dt^{2}}+\frac{d^{2}%
(\delta{\mathbf{y}}_{a})}{dt^{2}}\right]  +\frac{\Lambda m_{a}(\bar
{{\mathbf{x}}}_{a}+\delta{\mathbf{y}}_{a})}{3}\\
&  =-\frac{\partial V_{a}(\bar{{\mathbf{x}}}_{a}+\delta{\mathbf{y}}_{a}%
)}{\partial{\mathbf{x}}_{a}}=-\left[  \frac{\partial V_{a}(\bar{{\mathbf{x}}%
}_{a})}{\partial{\mathbf{x}}_{a}}+\frac{\partial^{2}V_{a}(\bar{{\mathbf{x}}%
}_{a})}{\partial{\mathbf{x}}_{a}\partial{\mathbf{x}}_{b}}\,\partial
{\mathbf{x}}_{b}\right]  .
\end{align}
where the potential $V_{a\Lambda}$ and its derivatives are
\begin{align}
V_{a\Lambda}  &  :=\sum_{a,i}\frac{\Lambda}{6}m_{a}(x_{a}^{i})^{2}:=\sum
_{a,i}\frac{\Lambda}{6}m_{a}(\bar{{\mathbf{x}}}_{a}+\delta{\mathbf{y}}_{a}%
^{i})^{2},\;\\
\frac{\partial V_{a\Lambda}}{\partial x_{ai}}  &  :=\sum_{a,i}\frac{\Lambda
}{3}m_{a}(\delta{\mathbf{y}}_{a}^{i}),\\
\frac{\partial^{2}V_{a\Lambda}}{\partial x_{ai}\partial x_{bj}}  &
=\frac{\partial}{\partial x_{bj}}\sum\frac{\Lambda}{3}m_{a}(\delta{\mathbf{y}%
}_{a}^{{\mathbf{i}}})=\frac{\Lambda}{3}m_{a}\delta_{ab}\delta_{ij}.
\end{align}
Cancelling the background terms, the perturbation equation is
\[
m_{a}\ddot{\delta{\mathbf{y}}}_{a}=-\sum_{b\neq a}\frac{\partial^{2}V_{a}%
}{\partial{\mathbf{x}}_{a}{\partial{\mathbf{x}}_{b}}}(\bar{{\mathbf{x}}}%
_{1},\bar{{\mathbf{x}}}_{2},\dots\bar{{\mathbf{x}}}_{N}){\delta{\mathbf{y}}%
}_{b}+m_{a}\frac{\Lambda}{3}\delta{\mathbf{y}}_{a}\,.
\]
The symmetric linear operator acting on ${\mathbf{y}}_{a}$ is minus the
Hessian of $V=V_{a}^{grav}+V_{a\Lambda}$ , considered as a function on the
3N-dimensional configuration space evaluated on the background solution$.$

\subsubsection{Force form}

Using the expression $\mathbf{F}_{ab}=-\frac{Gm_{a}m_{b}}{|\mathbf{x}%
_{a}-\mathbf{x}_{b}|^{3}}(\mathbf{x}_{a}-\mathbf{x}_{b})$\ $+\frac{\Lambda
m_{a}\mathbf{x}_{a}}{3}$for the force between the particles at ${\mathbf{x}%
}_{a}$ and ${\mathbf{x}}_{b}$and proceeding as before gives

\begin{eqnarray}
\mathbf{F}_{ab}(\bar{{\mathbf{x}}}_{a}+\delta{\mathbf{y}}_{a})  &
=&-\frac{Gm_{a}m_{b}}{|(\bar{{\mathbf{x}}}_{a}+\delta{\mathbf{y}}_{a}%
)-(\bar{{\mathbf{x}}}_{b}+\delta{\mathbf{y}}_{b}\mathbf{)}|^{3}}%
((\bar{{\mathbf{x}}}_{a}+\delta{\mathbf{y}}_{a})-(\bar{{\mathbf{x}}}%
_{b}+\delta{\mathbf{y}}_{b}\mathbf{)})\nn\\ &+&\frac{\Lambda m_{a}\mathbf{\ }}{3}%
(\bar{{\mathbf{x}}}_{a}+\delta{\mathbf{y}}_{a})\nn\\
&=&-\frac{Gm_{a}m_{b}}{|{\mathbf{\bar{x}}}_{b}-{\mathbf{\bar{x}}}_{a}|^{3}%
}(\bar{{\mathbf{x}}}_{ab}+\delta{\mathbf{y}}_{ab})-\frac{\partial}%
{\partial\mathbf{x}_{a}}\left[  \frac{Gm_{a}m_{b}}{|\mathbf{x}_{b}%
-\mathbf{x}_{a}|^{3}}\right]  (\bar{{\mathbf{x}}}_{ab})(\delta{\mathbf{y}}%
_{a}\mathbf{)}\nn \\ &+&\frac{\Lambda m_{a}\mathbf{\ }}{3}(\bar{{\mathbf{x}}}%
_{a}+\delta{\mathbf{y}}_{a})\;\mathbf{+O(}\delta{\mathbf{y}}_{a})^{2}%
\end{eqnarray}
to first order, where the partial derivative $(\partial/\partial\mathbf{x}%
_{a})$ is taken keeping all the other positions $\mathbf{x}_{b}\,\,(b\neq a)$
constant. This gives
\begin{equation}
\delta{\mathbf{F}}_{ab}=\frac{Gm_{a}m_{b}}{|\bar{{\mathbf{x}}}_{ab}|^{5}%
}\left\{  \delta{\mathbf{y}}_{ba}|{\mathbf{\bar{x}}}_{ab}|^{2}-3({\mathbf{\bar
{x}}}_{ba})({\mathbf{\bar{x}}}_{ba})\centerdot\delta{\mathbf{y}}%
_{ba})\right\}  +\frac{\Lambda m_{a}\mathbf{\ }}{3}\delta{\mathbf{y}}%
_{a}\nonumber
\end{equation}
and so
\begin{eqnarray}
m_{a}(\delta{\mathbf{y}}_{a})\ddot{}&=&\sum_{b\neq a}\frac{Gm_{a}m_{b}%
}{|{\mathbf{\bar{x}}}_{a}-{\mathbf{\bar{x}}}_{b}|^{5}}\left\{  \delta
{\mathbf{y}}_{ba}|{\mathbf{\bar{x}}}_{a}-{\mathbf{\bar{x}}}_{b}|^{2}%
-3({\mathbf{\bar{x}}}_{b}-{\mathbf{\bar{x}}}_{a})(({\mathbf{\bar{x}}}%
_{b}-{\mathbf{\bar{x}}}_{a})\centerdot\delta{\mathbf{y}}_{ba})\right\}\nn\\
&+&\frac{\Lambda m_{a}\mathbf{\ }}{3}\delta{\mathbf{y}}_{a}%
\end{eqnarray}

\subsubsection{Background cosmology with cosmological constant}

As before, put in a homothetic factor and separate variables: using
(\ref{homothetic}),
%\begin{equation}
%\textbf{x}_a = S(t)\textbf{r}_a
%\end{equation}
%then
(\ref{Newtlamb}) becomes
\begin{equation}
m_{a}\mathbf{r}_{a}\frac{d^{2}S(t)}{dt^{2}}=-\sum_{b\neq a}Gm_{a}m_{b}%
\frac{S(t)(\mathbf{r}_{a}-\mathbf{r}_{b})}{S^{3}(t)|\mathbf{r}_{a}%
-\mathbf{r}_{b}|^{3}}+\frac{\Lambda S(t)m_{a}\mathbf{r}_{a}}{3}.
\label{Newtlamb1}%
\end{equation}
%
%that is
%\begin{equation}
%m_a \textbf{r}_aS^2(t) \frac{d^2 S(t)}{dt^2} = -\sum_{a\neq b} G m_a m_b\frac{(\textbf{r}_a-\textbf{r}_b)}{|\textbf{r}_a-\textbf{r}_b|^3} + \frac{\Lambda S^3(t) m_a \textbf{r}_a}{3}
%\end{equation}
The argument goes through as before. This gives the result
\begin{equation}
m_{a}\mathbf{r}_{a}S^{2}(t)\frac{d^{2}S(t)}{dt^{2}}=-G\tilde{M}m_{a}%
\mathbf{r}_{a}+\frac{\Lambda S^{3}(t)m_{a}\mathbf{r}_{a}}{3} \label{Newtlamb3}%
\end{equation}
with $\tilde{M}$ defined exactly as before by (\ref{Newt3}). This implies
%\begin{equation}
%S^2(t) \frac{d^2 S(t)}{dt^2} = -G\tilde{M}  + \frac{\Lambda S^3(t) }{3}
%\end{equation}
%which is
the Raychaudhuri equation with cosmological constant:
\begin{equation}
\frac{1}{S(t)}\frac{d^{2}S(t)}{dt^{2}}=-\frac{G\tilde{M}}{S^{3}(t)}%
+\frac{\Lambda}{3} \label{Newtlamb51}%
\end{equation}
where matter causes deceleration and $\Lambda$ an acceleration. To integrate
when $dS/dt\neq0$, multiply by $S(t)dS/dt$ to get the Friedmann equation
%\begin{equation}
%\frac{1}{2}\left[\frac{dS(t)}{dt}\right]^2 =   \frac{G\tilde{M}}{S(t)} + E + \frac{\Lambda S^2(t)}{6}
%\end{equation}%
\begin{equation}
\frac{1}{2}\left[  \frac{\dot{S}(t)}{S(t)}\right]  ^{2}=\frac{G\tilde{M}%
}{S^{3}(t)}+\frac{E}{S^{2}(t)}+\frac{\Lambda}{6}%
\end{equation}
where $E$ is a constant of integration$.$

\subsubsection{Perturbed cosmology with cosmological constant}

We again apply this general formalism to the homothetically expanding
background solution ${\mathbf{\bar{x}}}_{a}=S(t){\mathbf{\bar{r}}}%
_{a},\ {\mathbf{\bar{r}}}_{a}=const$ and define $\delta{\mathbf{y}}%
_{a}=S(t){\mathbf{S}}_{a}(t).$ Then%
\begin{eqnarray}
&&m_{a}\frac{d^{2}}{dt^{2}}(S(t){\mathbf{S}}_{a})=  \nn \\  &&\sum_{b\neq c}%
\frac{Gm_{a}m_{b}}{S^{5}(t)|{\mathbf{\bar{r}}}_{a}-{\mathbf{\bar{r}}}_{b}%
|^{5}}S^{3}(t)\left\{  ({\mathbf{S}}_{b}-{\mathbf{S}}_{a})|{\mathbf{\bar{r}}%
}_{a}-{\mathbf{\bar{r}}}_{b}|^{2}-3({\mathbf{\bar{r}}}_{b}-{\mathbf{\bar{r}}%
}_{a})\cdot\mathbf{S}_{ba})({\mathbf{\bar{r}}}_{b}-{\mathbf{\bar{r}}}%
_{a})\right\} \nonumber\\
& +&\frac{\Lambda m_{a}\mathbf{\ }}{3}S(t){\mathbf{S}}_{a}(t)
\end{eqnarray}
giving the cosmological perturbation equation%
\begin{equation}
S^{2}m_{a}\frac{d^{2}}{dt^{2}}(S{\mathbf{S}}_{a})=\sum_{b\neq c}\frac
{Gm_{a}m_{b}}{|{\mathbf{\bar{r}}}_{a}-{\mathbf{\bar{r}}}_{b}|^{5}}\left\{
\bar{r}_{ba}^{2}{\mathbf{S}}_{ba}-3({\mathbf{\bar{r}}}_{ba}\cdot{\mathbf{S}%
}_{ba}){\mathbf{\bar{r}}}_{ba}\right\}  +\frac{\Lambda m_{a}\mathbf{\ }}%
{3}S^{3}(t){\mathbf{S}}_{a}(t)
\end{equation}
for perturbations with $\Lambda \neq 0$.

\subsubsection{Asymptotic solution}

Multiply by $(1/m_{a}S^{2})$, the growth of perturbations is given by
\begin{equation}
\frac{d^{2}}{dt^{2}}(S\,{\mathbf{S}}_{a})=\frac{1}{S^{2}}\sum_{b\neq c}%
\frac{Gm_{b}}{\bar{r}_{ab}^{5}}\left\{  \bar{r}_{ba}^{2}{\mathbf{S}}%
_{ba}-3({\mathbf{\bar{r}}}_{ba}\cdot{\mathbf{S}}_{ba}){\mathbf{\bar{r}}}%
_{ba}\right\}  +\frac{\Lambda\mathbf{\ }}{3}S(t){\mathbf{S}}_{a}(t)
\end{equation}
The first term on the right hand side goes to zero as $S\rightarrow\infty$.
Thus at late times
\begin{equation}
\frac{d^{2}}{dt^{2}}(S\,{\mathbf{S}}_{a})=\frac{\Lambda\mathbf{\ }}%
{3}(S{\mathbf{S}}_{a}).
\end{equation}
Assuming $\Lambda>0,$ this implies
\[
{\mathbf{S}}_{a}=\frac{\mathbf{S}_{0}\exp\sqrt{\frac{\Lambda\mathbf{\ }}{3}%
}(t-t_{0})}{S(t)}.
\]
where $\mathbf{S}_{0}$ is a constant vector. This means the density
perturbation is rapidly decreasing, as the exponential wins at late times:%

\[
\frac{d{\mathbf{S}}_{a}}{dt}=\frac{\mathbf{S}_{0}}{S(t)}\left[  \sqrt
{\frac{\Lambda\mathbf{\ }}{3}}-\frac{1}{S(t)}\frac{dS}{dt}\right]  \exp\left(
\sqrt{\frac{\Lambda\mathbf{\ }}{3}}(t-t_{0})\right)
\]
which changes sign when $\allowbreak\frac{1}{S(t)}\frac{dS}{dt}=\sqrt
{\frac{\Lambda\mathbf{\ }}{3}}.$This is when the vacuum energy wins over the
gravitational attraction, and structure formation ceases.

\section{The Dmitriev-Zel'dovich equations}

\label{sec:DZ}

We turn now to a different approach to deriving perturbation equations,based
in work of Dimitriev and Zeldovich, that is useful in n-body simulations
\cite{Bag05}.\medskip

One can group particles together to get identified subgroups, and coarse grain
to get equations for each subgroup. Then one can assume one subgroup - say a
system of galaxies - has little influence on the rest of the universe, which
is much larger; so this system moves in the averaged field of the background
universe, which is unaffected by its presence. In the case of just one
subgroup, this gives the Dimitriev-Zeldovich equations from Newton's equations
of motion, which are valid even when the situation is non-linear. This is the subject of sections (4.1) and (4.2). \newline

The Dmitriev-Zel'dovich equations contain the scale
factor $S(t)$ and are thus  time dependent. They nevertheless admit
a Lagrangian description (discussed in section 4.3)  
 and as a consequence satisfy the   conservation
of momentum and angular momentum by virtue of the translation and rotation
invariance of the Lagrangian, although the  expressions  
for the momentum and angular momentum in terms of position 
and velocities are time dependent because they contain the scale 
factor $S(t)$. Because of the time dependence, 
energy is no longer conserved, and as we discuss in section (4.4) 
the usual Virial Theorem takes
a modified form which is widely used in large scale structure studies.

The background Newtonian universe we are considering
is not invariant under Galilean  boosts and thus    
may be said to exhibit the spontaneous breakdown of Galilean invariance
just  as its relativistic version, the Friedmann-Lemaitre-Robertson-Walker
metric exhibits the spontaneous breakdown of Lorentz invariance.
Nevertheless, there remains a remnant of Galilean invariance
in the Dmitriev-Zel'dovich equations, which exhibit a form of the 
relativity principle which has some relevance for discussions 
of whether space is relative or absolute. This is discussed in section (4.5).
      
In section (4.6) we discuss the two-body problem 
according to the  Dmitriev-Zel'dovich equations and show 
how, in the adiabatic approximation, the orbits of planets around the sun
or stars around the galaxy participate in the general expansion
of the universe.

\subsection{Coarse Graining}

We start with the exact equations of motion for a large but finite number of
particles:
\begin{equation}
m_{a}\ddot{{\mathbf{x}}}_{a}=\sum_{b\neq a}\frac{Gm_{a}m_{b}({\mathbf{x}}%
_{b}-{\mathbf{x}}_{a})}{|{\mathbf{x}}_{a}-{\mathbf{x}}_{b}|^{3}}\label{dz1}%
\end{equation}
and assume that the particles fall into two classes, with a$=i,j,k...$ and
$a=I,J,K,....$ The second set form a cosmological background and we make the
approximation that their motion is unaffected by the first class of particles,
galaxies, whose motion is however affected both by the background particles
and their mutual attractions. Thus the equations of motion (\ref{dz1}) split
into two sets
\begin{equation}
m_{I}\ddot{{\mathbf{x}}}_{I}=\sum_{J\neq I}\frac{Gm_{I}m_{J}({\mathbf{x}}%
_{J}-{\mathbf{x}}_{I})}{|{\mathbf{x}}_{J}-{\mathbf{x}}_{I}|^{3}}\label{dz11}%
\end{equation}
for the background model and
\begin{equation}
m_{i}\ddot{{\mathbf{x}}}_{i}=\sum_{j\neq i}\frac{Gm_{i}m_{j}({\mathbf{x}}%
_{j}-{\mathbf{x}}_{i})}{|{\mathbf{x}}_{j}-{\mathbf{x}}_{i}|^{3}}+\sum_{J}%
\frac{Gm_{i}m_{J}({\mathbf{x}}_{J}-{\mathbf{x}}_{i})}{|{\mathbf{x}}%
_{J}-{\mathbf{x}}_{i}|^{3}}\label{dz2}%
\end{equation}
for the subgroup. We now assume that the background particles move
isometrically:
\begin{equation}
{\mathbf{x}}_{I}=S(t){\mathbf{r}}_{I}.\label{cen}%
\end{equation}
Then by the above argument, they must form a central configuration and $S(t)$
obeys the Friedmann equation (\ref{Fried12}).The deviation of the first set of
particles from this mean Hubble flow is given by
\begin{equation}
m_{i}\ddot{{\mathbf{x}}}_{i}=\sum_{j\neq i}\frac{Gm_{i}m_{j}(\mathbf{x}%
_{j}-{\mathbf{x}}_{i})}{|{\mathbf{x}}_{i}-{\mathbf{x}}_{j}|^{3}}+\sum_{J}%
\frac{Gm_{i}m_{J}S(t)({\mathbf{r}}_{J}-{\mathbf{r}}_{i})}{|S(t)({\mathbf{r}%
}_{J}-{\mathbf{r}}_{i})|^{3}}\label{dz22}%
\end{equation}
We replace the absolute positions of the galaxies by the conformally scaled
positions ${\mathbf{x}}_{i}=S(t){\mathbf{r}}_{i}(t)$ and obtain
\begin{eqnarray}
m_{i}\left(  S(t)\ddot{{\mathbf{r}}}_{i}+2\dot{S}(t)\dot{{\mathbf{r}}}
_{i}+\ddot{S}(t)\dot{{\mathbf{r}}}_{i}\right)  
&=& \frac{1}{S^2(t)} \sum_{j\neq i}
\frac{Gm_{i}m_{j}({\mathbf{r}}_{j}-{\mathbf{r}}_{i})}
{|{\mathbf{r}}_{i} -{\mathbf{r}}_{j}|^{3}} \nonumber\\
&+& \frac{1}{S^2(t)}
\sum_{J} \frac{Gm_{i} m_{J}({\mathbf{r}}_{J}-{\mathbf{r}}_{i})}
{|{\mathbf{r}}_{J}-{\mathbf{r}}_{i}|^{3} }\,.\label{dz3}
\end{eqnarray}
The second term on the right hand side of (\ref{dz3}) is the force $F_{i}$
exerted on the $i$th galaxies by the background particles. The numerical work
in \cite{BatGibSut03} provided very good evidence that for a large number of
background particles, the central configuration is to a very good
approximation statistically spherically symmetric and homogeneous. It follows
that the force exerted by the background is radial
\begin{equation}
\frac{1}{S^{2}(t)}\sum_{J}\frac{Gm_{i}m_{J}({\mathbf{r}}_{J}-{\mathbf{r}}_{i})}
{|{\mathbf{r}}_{J}-{\mathbf{r}}_{i}|^{3}}=-G\tilde{M}m_{i}{\mathbf{r}%
}_{i},\label{dx4}%
\end{equation}
where by (\ref{eq:ray}),
\begin{equation}
S^{2}\ddot{S}=-G\tilde{M}.\label{dx5}%
\end{equation}
Then the force term $F_{i}:=\frac{1}{S^{2}(t)}\sum_{J}\frac{Gm_{i}%
m_{J}({\mathbf{r}}_{J}-{\mathbf{r}}_{i})}{|{\mathbf{r}}_{J}-{\mathbf{r}}%
_{i}|^{3}}$ on the right hand side of (\ref{dz3}) cancels the third term on
the left hand side. We are left with
\begin{equation}
m_i\left(  S(t)\ddot{{\mathbf{r}}}_{i}+2\dot{S}(t)\dot{{\mathbf{r}}}%
_{i}\right)  =\frac{1}{S^{2}(t)}\sum_{j\neq i}\frac{Gm_{i}m_{j}({\mathbf{r}%
}_{j}-{\mathbf{r}}_{i})}{|\mathbf{r}_{i}-{\mathbf{r}}_{j}|^{3}},\label{dz6}%
\end{equation}
that is
\begin{equation}
\frac{d (S^2(t){\dot {\mathbf{r}}}_i )}{dt} 
=\frac{1}{S(t)}
\sum_{j\neq i} \frac{Gm_j({\mathbf{r}}_j-{\mathbf{r}}_i)}{|\mathbf{r}
_i-{\mathbf{r}}_j|^{3}}\label{dz7}
\end{equation}
which are  the Dmitriev-Zel'dovich equations \cite{Zel}.\newline

Writing this in terms of inertial coordinates 
$\mathbf{x}_i= S(t)\mathbf{r}_i$  rather than co-moving coordinates
the Dmitriev-Zel'dovich equation
takes the equivalent  form
\begin{equation}
{\ddot { \mathbf{x}}  }_i = \frac{ \ddot S}{S}  {\mathbf{x}}_i  +  \sum_{j\neq i} \frac{Gm_j({\mathbf{x}}_j-{\mathbf{x}}_i)}{|\mathbf{x}
_i-{\mathbf{x}}_j|^{3}} \label{alternativeeom} \,
\end{equation}
which appears in the work of \cite{Zhuk1,Zhuk2,Zhuk3}.

The equations of motion (\ref{dz7},\ref{alternativeeom}) 
contain the  time dependent scale factor $S(t)$ and its 
first (\ref{dz7}) or second (\ref{alternativeeom})
time derivative. nevertheless it is still possible to apply
the standard techniques of Lagrangian and Hamiltonian mechanics
as we shall show in the next subsection.

\subsection{Lagrangian version}

Peebles \cite{Pee89} has  shown that the Dmitriev-Zel'dovich equation
(\ref{dz7}) may derived from the (time-dependent) Lagrangian
\begin{equation}
L=\frac{1}{2}S^{2}\sum_{1\leq i\leq N}m_{i}{\dot{{\mathbf{r}}}}_{i}^{2}%
+\frac{1}{S}\sum_{1\leq i<j\leq N}\frac{Gm_{i}m_{j}}{|{\mathbf{r}}%
_{i}-{\mathbf{r}}_{j}|}=T-V\,, \label{Lagrangian} 
\end{equation}
with
\begin{equation}
T=\frac{1}{2}\sum_{1\leq j\leq N}S^{2}m_{i}{\dot{{\mathbf{r}}}}_{i}^{2}\,,
\end{equation}

\begin{equation}
V=-\frac{1}{S}\sum_{1\le i<j\le N}\frac{Gm_{i}m_{j}}{|{\mathbf{r}}%
_{i}-{\mathbf{r}}_{j}|}\,.
\end{equation}

The Lagrangian (\ref{Lagrangian}) differs from the Lagrangian  
\begin{equation}
\tilde L= \frac{1}{2}
\sum_{1\leq i\leq N}  \bigl ( m_{i} {\dot{\mathbf{x}} }_i^2 +
\frac{\ddot S}{S} m_i{\mathbf{x}}_i^2 \bigr)   
+\sum_{1\leq i<j\leq N}\frac{Gm_{i}m_{j}}{|{\mathbf{x}}%
_{i}-{\mathbf{x}}_{j}|} \,,
\end{equation}
where $\mathbf{r}_i= \frac{\mathbf{x}_i}{S(t)}$,  
by a total time derivative. Therefore it should  gives rise to the same
equations of motion. This is easily checked since
the Euler Lagrange equations of $\tilde L$  are 
in fact (\ref{alternativeeom}).

By virtue of translation and rotational invariance of $L$, the equations of
motion conserve total momentum
\begin{equation}
{\mathbf{P}}=\sum_{1\leq i\leq N}{{\mathbf{p}}}_{i}=S^2 \sum_{1\leq i\leq N}%
m_{i}\dot{{\mathbf{r}}}_{i} \Rightarrow d{\mathbf{P}}/dt =0,  
\end{equation}
and total angular momentum
\begin{equation}
{\mathbf{L}}=\sum_{1\leq i\leq N}{\mathbf{r}}_{i}\times{\mathbf{p}}_{i} \Rightarrow d{\mathbf{L}}/dt = 0
\end{equation}
with
\begin{equation}
{\mathbf{p}}_{i}=\frac{\partial L}{\partial{\dot{{\mathbf{r}}}_{i}}}%
=\frac{\partial T}{\partial{\dot{{\mathbf{r}}}}_{i}} =
 S^2 m_{i} \dot{\mathbf{r}}_{i} \,.
\end{equation}

\subsection{Energy theorem and virial theorem}

Because of the time dependence of the Lagrangian, 
the energy or Hamiltonian $H$ is not
conserved. For a general Lagrangian system we have
\begin{equation}
\frac{dH}{dt}=-\frac{\partial L}{\partial t}%
\end{equation}
where
\begin{equation}
H=\sum_{1\le i\le N}{\mathbf{p}}_{i}\cdot{\dot{{\mathbf{r}}}}_{i}-L
\end{equation}
In our case
\begin{eqnarray}
H&=&T+V \\ &=&  
 \frac{1}{ S^2(t) }
\sum_{1\leq j\leq N}  \frac{ \mathbf{p}^2 _i} {2m_i} -\frac{1}{S}\sum_{1\le i<j\le N}\frac{Gm_{i}m_{j}}{|{\mathbf{r}}%
_{i}-{\mathbf{r}}_{j}|}\,.
\end{eqnarray}
and we have the so-called \textit{Cosmic Energy Theorem} 
\cite{Irvine,Layzer, Zel}
%\footnote{I suppose the correct citations
%would be \cite{Layzer,Zel,Irvine} and possibly Irvine's thesis}.%
\begin{equation}
\frac{dH}{dt}=\frac{\dot{S}}{a}\bigl (2T-V\bigr )
\end{equation}
Note that if $V$ can be neglected, or for freely moving non-relativistic
particles, the energy is pure kinetic and redshifts as $\frac{1}{S^{2}}$.

One may easily extend this result to include a cosmological term or possible
dark energy effects cf. \cite{Shtanov:2010iy} or \cite{Avelino:2013ira}.

To obtain the so-called \textit{Cosmic Virial Theorem} \cite{Pee76} we recall
that for a general Lagrangian system%

\begin{align}
\frac{d}{dt}\sum_{1\le i\le N}{\mathbf{p}}_{i}\cdot{{\mathbf{r}}}_{i}  &
=\sum_{1\le i\le N}\dot{{\mathbf{p}}}_{i}\cdot{\dot{{\mathbf{r}}}}_{i}%
+\sum_{1\le i\le N}\dot{{\mathbf{p}}}_{i}{{\mathbf{r}}}_{i}\\
&  =H+L+\sum_{1\le i\le N}{\dot{{\mathbf{r}}}}_{i}\cdot{\frac{\partial
L}{\partial{{\mathbf{r}}}}_{i}}%
\end{align}

In our case we get
\begin{equation}
\frac{d}{dt}\bigl (S^{2}\frac{dI}{dt}\bigr )=2T+V
\end{equation}
where
\begin{equation}
I=\frac{1}{2}\sum_{1\le i\le N}m_{i}{{\mathbf{r}}}_{i}^{2}\,.
\end{equation}
If we time average and assume that the average of the rhs is zero
we get 
\begin{equation}\label{eq:vir}
<2T+V> =0 \Rightarrow %\frac{d}{dt}\bigl (S^{2}\frac{dI}{dt}\bigr )=0  \Rightarrow  
S^{2}<\frac{dI}{dt}>=const  
\end{equation} which is the standard result (see \cite{First}).
This will be true when the local system has decoupled from the cosmic expansion; 
otherwise we get
\begin{equation}
  (S^2 <dI/dt>|_{t=t_1} - S^2 <dI/dt>|_{t=t_2})= \int_{t_1}^{t_2}(<2T+V>)dt 
\end{equation}
which will be non-zero for systems coupled to the cosmic  expansion. 
Conditions  for the Virial Theorem condition on the left of  (\ref{eq:vir}) to hold are given in \cite{Poll64}. In essence the result holds because the asymptotic average of the derivative of a bounded function is necessarily zero, thus it will hold for any bound system of self-gravitating particles.\footnote{See http://www.mathpages.com/home/kmath572/kmath572.htm for more details.}

\subsection{Galilean Invariance}

The equations of motion are invariant under the generalised Galilean
transformations
\begin{equation}
{{\mathbf{r}}}_{i}\rightarrow{{\mathbf{r}}}_{i}+{\mathbf{a}(t)}\label{Gal}%
\end{equation}
where
\begin{equation}
\frac{d\left(  S^{2}(t){\dot{{\mathbf{a}}}}\right)  }{dt}=0\,.
\end{equation}

The Lagrangian itself is not invariant under (\ref{Gal}) but changes by a time
derivative. We also have that that the centre of mass ${\mathbf{R}}$, defined
by
\begin{equation}
{\mathbf{R}}=\frac{1}{M}\sum_{1\le i\le N}m_{i}{{\mathbf{r}}}_{i}\,,
\end{equation}
moves as
\begin{equation}
\frac{d\left(  S^{2}(t){\dot{{\mathbf{R}}}}\right)  }{dt}=0\,.
\end{equation}
and that by means of a generalised Galilean transformation of the form
(\ref{Gal}) we may pass to Barycentric coordinates for which ${\mathbf{R}}=0$.

It is well known that while  Leibnitz adhered to a relational theory
of space, i.e. that absolute positions are unobservable, Newton
appears to his critics  at least to have favoured  the idea that  
space is absolute or perhaps more accurately, 
 that something (God?) determines an absolute standard
of rest.   
In the late nineteenth century, by which time the 
proper definition and consequent  arbitrariness 
of an inertial frame was finally understood 
\cite{Mach,ThomsonTait,Thomson1,Thomson2,Lange,Muirhead}, 
there were also suggestions \cite{ThomsonTait} that despite the fact that
fundamental
laws of dynamics were Galilei invariant, 
a privileged inertial frame, sometimes called the \lq Body Alpha \rq
\cite{Neumann} 
might be identified with the rest frame of all of the particles 
in the universe, assumed finite and that may always refer the fundamental
equations
of dynamics to that frame. As noted above, from a modern
perspective, according to  which our background universe spontaneously breaks
Galilei invariance with a cosmological rest frame defined by the cosmic background radiation, the puzzle is why the motion of bodies
within it should exhibit an albeit modified form of Galilei invariance.
The answer we see at the Newtonian level is that it is inherited from the underlying Galilei invariance of the equations (\ref{Newt})
which  we started with. 

From a more practical viewpoint it is worth perhaps worth  remarking  
that the 2nd realisation of the  International Celestial Reference System 
(ICRF2) uses 3414 extragalactic radio sources observed by Very Long Baseline
Interferometry (VLBI) each of whose motion is presumably  governed 
by the Dmitriev-Zeldovich equations.   

\subsection{Two-Body Problem for the Dmitriev-Zeldovich equations }

The Dmitriev-Zeldovich equations for two bodies may be obtained from the
Lagrangian
\begin{equation}
L={\frac{1}{2}}S^{2}(t)(m_{1}\dot{\mathbf{r}}_{1}^{2}+m_{2}\dot{\mathbf{r}%
}_{2}^{2})+\frac{Gm_{1}m_{2}}{S(t)|\mathbf{r}_{1}-\mathbf{r}_{2}|}%
\end{equation}
This may be re-arranged to give give
\begin{equation}
L={\frac{1}{2}}S^{2}(t)M(\frac{m_{1}\dot{\mathbf{r}}_{1}+m_{2}\dot{\mathbf{r}%
}_{2}}{m_{1}+m_{2}})^{2}+{\frac{1}{2}}S^{2}(t)\mu\dot{\mathbf{r}}^{2}+\frac
{1}{S(t)}\frac{GM\mu}{|\mathbf{r}|}%
\end{equation}
where $M=m_{1}+m_{2},\,\mu=\frac{m_{1}m_{2}}{m_{1}+m_{2}},\,\mathbf{r}%
=\mathbf{r}_{1}-\mathbf{r}_{2}$. The first term give the motion of the centre
of mass $\mathbf{R}=\frac{m_{1}\mathbf{r}_{1}+m_{2}\mathbf{r}_{2}}{m_{1}%
+m_{2}}$ and the second and third terms the relative motion. If
$S(t)=\mathrm{constant}$ this is the standard Kepler problem. If $S(t)$ varies
with time, except for the special Vinti-Lynden-Bell case in which the solution
may be expressed in terms of the solution of the time independent case (see
\cite{Vinti,Lyn,Duval}), we must resort to an approximation. However relative
angular momentum $\mathbf{L}$ is conserved and by rotational invariance we may
still reduce the problem to one in the equatorial plane orthogonal to
$\mathbf{L}$.

\subsubsection{Adiabatic Invariants}

The standard approach to problems of this kind
is to find the adiabatic invariants of the
time-independent motion and then, for slowly $S(t)$ they should be constant.
The relevant adiabatic invariants are $\frac{1}{2\pi}\oint p_{r}dr$ and
$\frac{1}{2\pi}\oint p_{\phi}d\phi$.

The motion with $S(t)$ constant is an ellipse with semi-major axis $a$ and
eccentricity $\epsilon$ and semi-major axis $b=a\sqrt{1-\epsilon^{2}}$ and
semi-latus rectum $l=a(1-\epsilon^{2})$.
\begin{equation}
\frac{1}{r}=\frac{1+\epsilon\cos\phi}{a(1-\epsilon^{2})}\mathbf{r}%
\end{equation}
and so we want to know how $\epsilon$ and $a$ vary with time in the
adiabatic approximation. We have
\begin{equation}
\frac{1}{2\pi}\oint p_{r}dr=p_{\phi}\Bigl (\frac{1}{\sqrt{1-\epsilon^{2}}%
}-1\Bigr )\,. \label{adiabtic}
\end{equation}

An illuminating derivation of (\ref{adiabtic})
 is to be found in 
\cite{TaitSteele,Original} as follows. Let $p$ and $p^{\prime}$ be the
perpendicular distances from the foci $F$ and $F^{\prime}$ of an ellipse to
the tangent at the point $P$ whose focal distances are $r$ and $r^{\prime}$.
Since the two focal radii are equally inclined to the tangent we have
\begin{equation}
\frac{r}{p}=\frac{r^{\prime}}{r^{\prime}}\,.
\end{equation}
Now the pedal equations of the ellipse, respect to the foci are
\begin{equation}
\frac{l}{p^{2}}=\frac{2}{r}-\frac{1}{a}\,,\qquad\frac{l}{{p^{\prime}}^{2}%
}=\frac{2}{r^{\prime}}-\frac{1}{a}\,,
\end{equation}
where $l=a(1-\epsilon^{2})$ is the semi-latus rectum and $a$ the semi-major
axis, and $\epsilon$ its eccentricity. Thus
\begin{equation}
\frac{lr^{\prime}}{pp^{\prime}}=2-\frac{r}{a}\,,\qquad\frac{lr}{pp^{\prime}%
}=2-\frac{r^{\prime}}{a}\,,\qquad\,
\end{equation}
Addition yields
\begin{equation}
\frac{l(r+r^{\prime})}{pp^{\prime}}=2-\frac{r+r^{\prime}}{a}\,.
\end{equation}
But since $r+r^{\prime}=2a$, it follows that
\begin{equation}
pp^{\prime}=b^{2}\,,
\end{equation}
\textbf{r}where $b=a\sqrt{1-\epsilon^{2}}$ is the semi-minor axis.

Now consider a particle moving in an elliptic orbit about the focus $F$. It is
well known that Kepler's third law states that area $A$ swept out by the
radius vector from the focus $F$ is proportional to the time:
\begin{equation}
\int dA={\frac{1}{2}}\int pds={\frac{1}{2}}\int pvdt={\frac{1}{2}}h\int dt\,,
\end{equation}
where $h=pv=\frac{1}{m}p_{\phi}$ is the angular momentum per unit mass.

Less well known is the fact \cite{TaitSteele,Original} that the area
$A^{\prime}$ swept out by the radius vector from the focus $F^{\prime}$ is
proportional to the action:
\begin{equation}
\int dA^{\prime}={\frac{1}{2}}\int p^{\prime}ds={\frac{1}{2}}\int\frac{b^{2}%
}{p}ds={\frac{1}{2}}\frac{b^{2}}{h}\int vds=\frac{b^{2}}{2p_{\phi}}%
\int(p_{\phi}d\phi+p_{r}dr)\,.
\end{equation}
Now for one complete circuit $\oint dA=\oint dA^{\prime}=\pi ab$ and hence
\begin{equation}
\frac{1}{2\pi}\oint(p_{\phi}d\phi+p_{r}dr)=\frac{p_{\phi}}{\sqrt
{1-\epsilon^{2}}}\,.
\end{equation}
We deduce that the eccentricity $\epsilon$ is independent of time in the
adiabatic approximation.

We also have
\begin{equation}
\frac{1}{a(1-\epsilon^{2})}=\frac{\mu
SGm_{1}m_{2}}{p_{\phi}^{2}}\,,
\end{equation}
and we deduce that
\begin{equation}
S(t)a=\mathrm{constant}\,.
\end{equation}
In other words the size of the orbit in inertial coordinates $\mathbf{x}%
=S(t)\mathbf{r}$ is independent of time.

\subsubsection{The effect of the expansion of the universe on the solar
system}

Expressed in another way, one may use the size of binary systems as a
\textquotedblleft ruler\textquotedblright\ with which to measure the
\textquotedblleft expansion of the universe\textquotedblright. This is
consistent with the analysis of the effect of the Slipher-Hubble expansion on
the solar system \cite{MatGiu06}.

Since in our Newtonian model the distance to the  the galaxies
$|\bx |$ is increasing in accordance with the Slipher-Hubble law,
we see that the solar system has, to the approximation we are working to, 
a  fixed size relative to  which the universe  may be said to be expanding. 
One may compare this situation with the well known   Einstein-Strauss
or Swiss cheese model in General Relativity (see next sction for
the Newtonian version of this).   Each vacuole. i.e. spherical hole
in the cheese,   is occupied
by a locally static Scwharzschild solution. 
The boundary of the vacuole moves radially outwards with respect to the the static  Schwarzschild solution  with the same motion as a freely moving 
radial geodesic.
Within the vacuole one may imagine a test particle moving on circular
geodesic of constant radius.  Clearly the boundary of the vacuole
is expanding relative to this circular orbit. Thus our Newtonian
result, based on the theory of adiabatic invariants,   
is perfecly consistent with what one obtains according to general
relativity.

\section{Clustering and Swiss Cheese}

Most of the work in \cite{BatGibSut03} was concerned with the all when all
masses $m_{a}$ were taken to be equal. The resulting distributions, for
$N\approx10^{4}$ were extremely homogeneous, resembling a random close packing
of spheres, with the masses located at the centres of the spheres and with the
mean separation mentioned earlier. Interestingly the introduction of a
particle with very much larger mass, had the effect of evacuating a much
larger sphere, the mass again being located at the centre of the large
vacuole, the average density being maintained. This is the Newtonian analogue
of Einstein and Strauss's Swiss Cheese model in General Relativity
\cite{EinStr}. It is not what we will later consider as a void.

No evidence was found for clustering or hierarchical structure in the central
configurations investigated in \cite{BatGibSut03} and this appears to be
consistent with the results of \cite{Buc90} on the absence of clustering in
central configurations.

\subsection{Negative mass and the motion of voids}

We shall take (\ref{dz7}) as the equation of motion governing the interaction
of voids and regions of over density (``attractors''). It has the form of
Newtons' law ( with respect to the time $\tau$) but for which the effectively
Newton's $S(\tau)G$ varies with time $\tau$. If $S(t)G\propto{\frac{1}{\tau}}$
one may, by redefining the time variable, reduce the problem to a time
independent Newton's constant \cite{Vinti,Lyn,Duval}. Unfortunately this is
not possible in our case and we have to consider a genuinely time-dependent
Newton's constant. Despite that we can deduce that

\begin{itemize}
\item Both voids and attractors fall in the same way in a gravitational field,
that is their inertial masses and passive gravitational masses are equal.

\item Attractors attract and voids repel. Thus attractors have positive active
gravitational mass and voids have negative active gravitational mass. Thus
both attractors and voids are attracted towards attractors \textsl{and both
are repelled from from voids}. The direction they actually move of course
depends on their initial velocities.

\item Attractors have positive inertial masses and positive passive
gravitational masses, voids have negative inertial masses and negative passive
gravitational masses.

\item Action and reaction are equal and opposite and so the centre of mass
moves with constant velocity and angular momentum is conserved.
\end{itemize}

Counter-intuitive motion of this sort appears to have first been contemplated
by F\"oppl \cite{Foppl1,Foppl2} before the advent of General Relativity. It is
in accordance with the behaviour predicted for general relativity by Bondi
\cite{Bondi3,Bondi4}. Bondi showed that despite the uniform motion of the
barycentre
\begin{equation}
\sum_{a}m_{a}\mathbf{x}_{a}\,. \label{barycentre}%
\end{equation}
this could lead to run away solutions. In fact for two bodies with
$m_{1}=-m_{2}$ and $m=|m_{1}|=|m_{2}|$, (\ref{barycentre}) is compatible with
constant separation
\begin{equation}
\mathbf{x}_{1}-\mathbf{x}_{2}=\mathbf{d}\,
\end{equation}
where $\mathbf{a}$ is a constant vector. The accelerations of both bodies are
given by given by
\begin{equation}
{\frac{aGm}{|\mathbf{d}|^{2}}}\,.
\end{equation}
In the case considered by Bondi \cite{Bondi3,Bondi4} the effective Newton's
constant was constant, and hence the mutual acceleration was constant. Bondi
succeeded in demonstrating the existence of exact solutions of Einstein's
equations exhibiting this effect and it was shown in \cite{Gibbonsmotion} that
negative mass naked singularities could chase regular positive mass black
holes (see also \cite{GibbonsHartnollIshibashi}). Gravitational repulsion due
to uncompensated voids has been pointed out previously by Piran \cite{Piran}. For other
studies of the gravitating properties of negative masses see
\cite{Treder2,Foppl}.

\section{Conclusion}

In this paper we have explored the extent to which
an analytic treatment of a  purely discrete  Newtonian particle model
can be useful in studying questions in  cosmology and large scale structure
formation. Our main tool has been what we have referred to as the
Dmitrive-Zeldovich equations \cite{Zel} which are widely used in numerical
simulations.  We have given a purely  Newtonian point particle  derivation.
There exist many other  approaches based in Newtonian 
fluid mechanics or a mixture of both fluid and particle viewpoints
e.g. \cite{Pee93}. Of course the equations can be obtained as a Newtonian
limit of General Relativity and such  a treatment may be found in  
\cite{Zhuk1,Zhuk2,Zhuk3}.

\end{document}